\documentclass[review,12pt]{elsarticle}

\usepackage{graphicx}
\usepackage{amsmath}
\usepackage{nicefrac}
\usepackage{soul} 
\usepackage{color}
\usepackage{nicefrac}
\usepackage{xcolor}
\usepackage{enumitem}
\usepackage[english]{babel}
\newcommand{\Ffs}{4f$_\text{s}$} 
\newcommand{\Sfs}{6f$_\text{s}$} 
\newcommand{\FfF}{4f$_\text{F}$} 
\newcommand{\Fft}{4f$_{\tau}$} 
\newcommand{\SfH}{6f$_\text{H}$} 
\newcommand{\SfHc}{6f$_\text{Hc}$}

\journal{Applied Ocean Research}

\begin{document}

\begin{frontmatter}



\title{Interaction of water surface waves with periodic and quasiperiodic cylinder arrays}


\author[inst1,inst3]{Joseph A. Smerdon\corref{cor1}}
\cortext[cor1]{}
\ead{jsmerdon@uclan.ac.uk}

\affiliation[inst1]{organization={Jeremiah Horrocks Institute for Mathematics, Physics and Astronomy, University of Central Lancashire},
            city={Preston},
            postcode={PR1 2HE}, 
            country={United Kingdom}}

\author[inst2]{Sam Coates}
\author[inst3]{Bogdan J. Matuszewski}
\author[inst4]{Benedict D. Rogers}

\affiliation[inst2]{organization={Department of Physics, University of Liverpool},
            city={Liverpool},
            postcode={L69 3BX}, 
            country={United Kingdom}}

\affiliation[inst3]{organization={School of Engineering and Computing, University of Central Lancashire},
            city={Preston},
            postcode={PR1 2HE}, 
            country={United Kingdom}}

\affiliation[inst4]{organization={School of Engineering,  University of Manchester},
            city={Manchester},
            postcode={M13 9PL}, 
            country={United Kingdom}}
\begin{abstract}
Inspired by transformation optics and photonic crystals, this paper presents a computational investigation into the interaction between water surface waves and array waveguides of cylinders with multiple previously unexplored lattice geometries, including, for the first time, quasiperiodic geometries. Extending beyond conventional square and hexagonal periodic arrays, transformation optics has opened up entirely new opportunities to investigate water wave propagation through arrays based on quasiperiodic lattices, and quasiperiodically arranged vacancy defects.  Using the linear potential flow open-source code Capytaine, missing element and $\tau$-scaled Fibonacci square lattices, the Penrose lattice, hexagonal $H_{00}$ lattice and Amman-Beenker lattice are investigated.  The existence of band gaps for all arrays is observed. A hexagonal lattice with vacancy defects transmits the least energy.  Bragg diffraction consistent with azimuthal rotational symmetry is observed from all arrays.  Bragg resonance causes reflection from arrays, resulting in multiple Bloch band gaps. Away from Bragg resonance, waves will distort significantly to achieve periodic relationships with arrays, supporting transformation-based waveguides.  The possible uses include adaptation to more versatile waveguides with applications such as offshore renewable energy and coastal defence.
\end{abstract}

\begin{graphicalabstract}
\includegraphics[width=\textwidth]{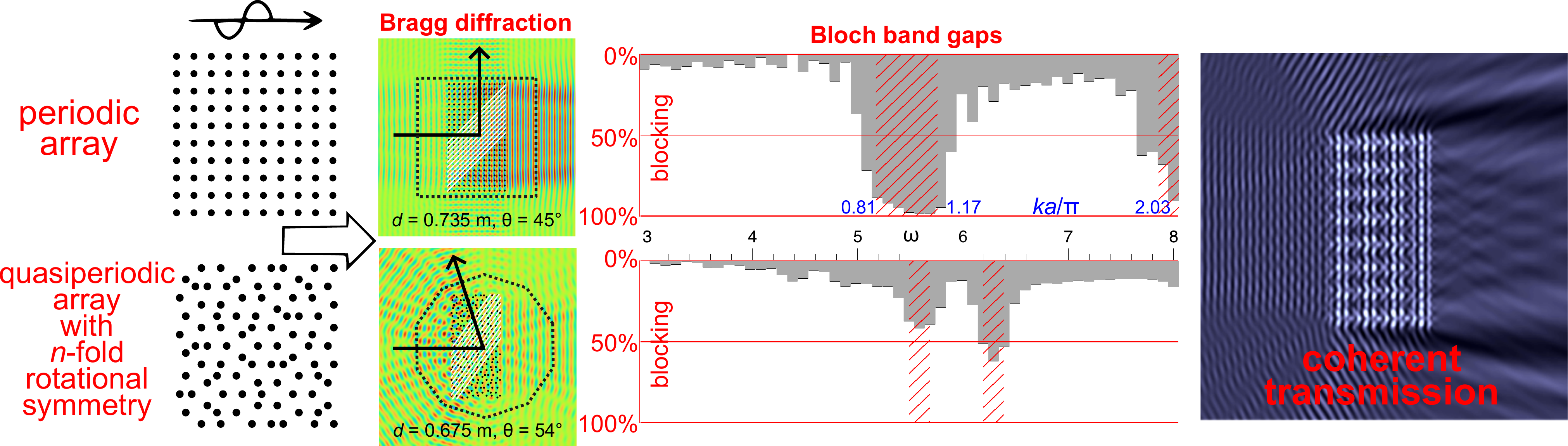}
\end{graphicalabstract}

\begin{highlights}
\item Inspired by transformation optics, quasiperiodic array water waveguides investigated
\item Propagation through arrays supports a transformation approach to waveguide design
\item Near 100\% redirection of wave energy achieved with minimal arrangements
\item Bragg resonance and Bloch band gaps in quasiperiodic lattices identified
\item Bragg diffraction consistent with rotational symmetry
\end{highlights}


\end{frontmatter}


\section{Introduction}

The vast potential of the revolution in optical waveguide engineering known as `transformation optics' has stimulated optics research for the past two decades \cite{pendry2006controlling,chen2010transformation}.  Combined with contemporary developments in manufacturing capability, particular success has been found in constructing photonic crystals: metamaterials with structures on length scales similar to the wavelength of the radiation. By modifying the permittivity and the permeability of the propagation medium, photonic crystals have been shown to exert deep influence on electromagnetic wave propagation \cite{joannopoulos1997photonic,dong2015conical}, leading to the demonstration of superlenses and invisibility devices \cite{pendry2006controlling}.  A similar field is transformation acoustics, in which the Young's modulus of the propagation medium is modified, with similarly useful results \cite{chen2010acoustic}. 

Inspired by transformation optics and photonic crystals, in this article we present an investigation into the interaction of water surface waves with `metamaterials' consisting of arrays of vertical cylinders in crystalline and other arrangements.

Improving our ability to manage and harness the energy in ocean waves is a problem of paramount importance, and has been the focus of much work over recent years \cite{zhu2024controlling,zhang2018observation,zhang2012band,wang2013experimental,wang2014focusing,hu2005refraction,hu2003complete,hu2004superlensing,tang2006omnidirectional,yang2009observation,linton1990interaction}. 
Free-surface water waves (disturbances of water) have been comprehensively studied analytically \cite{longuet1976deformation}, numerically \cite{Kirby2016}, and through observations both in controlled physical experiments and field measurements  \cite{Blenkinsopp_etal_2010,Blenkinsopp_etal_2012}. It has been shown that water surface waves share many of the same features as waves in other areas of physics (including fundamental processes such as reflection, refraction, diffraction, etc.). However, water waves also exhibit unique behaviour including shoaling (change of wave height) and breaking (collapse of waves) when they propagate into progressively shallower water. While the increasing emphasis on extracting energy from ocean waves and coastal protection has expanded our knowledge and understanding, new research is still needed, particularly to improve our understanding of complex wave interactions and the exploitation of wave energy devices.

There are several approaches to investigating the interaction of water waves with structures.  In the predominant approach, which started before simulation via computation became mainstream, efforts are focused on identifying those systems which lend themselves to analytical solutions.  The majority of these systems are 2-dimensional problems, $x$ being the propagation dimension and $y$ being the depth. The obstacles are modelled as variations in the propagation medium as functions of $x$ and $y$, including floating horizontal, vertical or otherwise mathematically defined bars, plates and 1-dimensional periodic structures \cite{peters1950effect,zhang2024resonant,liu2019bloch}, fixed versions of similar structures \cite{walker2005wave}, or similar structures applied to the sea bed \cite{xie2011analytical,linton2011water}.

Parallel behaviour in the interactions between water waves and periodic structures and between x-rays and crystals was noticed early on -- in particular, the advantageous Bragg resonance, in which waves are reflected from a periodic array and a Bloch frequency band gap is formed in which waves cannot propagate \cite{davies1982reflection,Mei_1985}.  In arrangements which are symmetrical in the propagation direction, zero-reflection wave modes, in which waves are able to propagate freely through the structure, are also possible \cite{xie2011analytical,goyal2025optimal}.  These phenomena in crystals underpin x-ray and electron diffraction, and the understanding of the behaviour of electrons in solids, including the band gap in semiconductors and conduction in metals  \cite{bloch1929quantenmechanik}.

Where analytical solutions were possible, they were developed, including of 3-dimensional systems of infinite 2-dimensional arrays of modulations of the propagation medium such as floating discs, based on square (4-fold) and triangular/hexagonal (6-fold) lattices \cite{chou1998band}. Arrays of surface-piercing cylinders were also investigated, with periodic boundary conditions (PBC) for infinite arrays \cite{mciver2000water,hu2005refraction,linton1990interaction} and without PBC for finite arrays \cite{liu2011propagation,ha2002propagation,ohl2001water}.  To the best of our knowledge, all infinite and finite periodic arrangements exhibit Bloch band gaps. The relationship between a 2-dimensional array and the propagation direction, however, is no longer implicit; waves with different propagation directions experience different lattice periodicities and different transmission properties, including band gap. The variability in periodicity is a function of the $n$-fold rotational symmetry of the lattice on which the array is based.

In addition to periodic lattices based on crystalline symmetries, \textit{quasiperiodic} lattices may also be constructed that have no translational symmetry but higher-order rotational symmetries (higher values of $n$), such as the pentagonal Penrose tiling (10-fold) \cite{penrose1979pentaplexity} and the octagonal Ammann-Beenker tiling (8-fold) \cite{arnouxtilings}.  Quasicrystals (quasiperiodic crystals) based on these lattices are the appropriate analogue of periodic crystals \cite{shechtman1984metallic,levine1984quasicrystals}. The higher orders of rotational symmetry available in quasiperiodic structures are attractive because they offer the same behaviours that periodic structures do, such as band gaps, but with greater isotropy and hence less angular selectivity \cite{rechtsman2008optimized}, supporting, in principle, manipulation of waves travelling in different directions.  Quasicrystalline metamaterials have been shown to have fractal transmission spectra with several large gaps \cite{davies2023graded}. 

Regarding the comparison with transformation optics (in which the relative permittivity and permeability are modified), the analogous properties which affect propagation of water waves are gravity and depth.  Despite the clear analogies between transformation optics and water wave propagation, such lattice arrangements of cylinder arrays and scatterers are yet to be fully investigated for water waves.

With respect to water surface waves, and in the long-wavelength regime ($\lambda\gg a$, where $\lambda$ is the wavelength and $a$ is a characteristic crystal dimension, defined as the lattice parameter for periodic arrays), it has been shown that the presence of an array of vertical cylinders alters the refractive index via modification of the effective gravity within the array, allowing construction of lenses and prisms \cite{hu2005refraction}.  This approach has also been shown experimentally to result in superlensing \cite{hu2004superlensing}. Moreover, analogous to the manipulation of permittivity/permeability in photonic materials, manipulating the depth via modification of either the floor or the free surface can lead to the full range of transformation optics effects \cite{wu2018double,qin2023superscattering,wang2015manipulating}.


Generally, modifying the density of a lattice of elements of given permeability/permittivity [Young's modulus] causes optical [acoustic] waves to propagate in the direction of the density modifications. The situation for water surface waves is essentially the same.  

A water surface waveguide array can modify the effective gravity \textit{and} effective depth in the region of the array.  This is in contrast to modifications of bathymetry, which allow control only over the effective depth, and which, in the context of coastal defence, are expensive and temporary.

Regarding analytical approaches, as mentioned, the application of PBC allows the derivation of transmission functions for linear waves through arrays \cite{Kakuno}. If the problem is restricted to two dimensions, as is possible for infinite depth or floor-mounted cylinders, it is possible to derive transmission functions even for finite arrays \cite{linton1990interaction}.  However, as suggested by the name, PBC cannot describe quasiperiodic geometries of any kind. In principle PBC may be applied to higher-dimensional space, from which a quasiperiodic two- or three-dimensional solution may be obtained (using the \textit{cut-and-project} method \cite{de1974pseudo,Elser}). Such a solution preserves the general behaviour expected from an infinite quasiperiodic lattice, but interpreting the implications in real space is non-trivial and a source of unwanted complexity.

Regarding computational approaches, any conceivable system may be investigated using fully-non-linear potential flow (FNLPF) calculations, but these calculations are extremely expensive in terms of time and computing resources.  Alternatively, the popular boundary-element-method (BEM) may be used to investigate arbitrary three-dimensional problems within the fully linear regime (including those that can be simplified with PBC, with concomitant increases in efficiency). 

Here, we demonstrate the use of the Capytaine \cite{ancellin_capytaine_2019} BEM python package to investigate a three-dimensional problem: finite arrays of finite-length cylinders in infinite-depth water. This problem, despite its apparent simplicity, is intractable via any other method except FNLPF calculations, and is thus ideal to demonstrate the validity of the BEM method to study arrays with arbitrary properties. 

The robustness of the approach with respect to arbitrary arrays is of great importance for the development of the field of water wave array waveguides.  The versatility of the approach enables the study of an arbitrary number of optimization parameters and thus, in principle, the tailoring of a waveguide to a particular situation.

We explore the interaction between monochromatic water surface waves and small ($N<300$) arrays of vertical surface-piercing cylinders.  These cylinders are arranged in translational and rotational symmetries previously uninvestigated and, in particular, we measure their capacity to allow or stop wave propagation in the near-Bragg regime ($\lambda\approx 2a$). Our focus is on the detailed structure of the arrays, which include periodic, quasiperiodic, and defected periodic geometries with the resulting effects on water surface wave propagation. 

The paper is structured as follows: first, we introduce the arrays that are investigated, and characterise them in terms of their symmetries. Then, we describe the wave modelling and numerical setup using the open-source software Capytaine \cite{ancellin_capytaine_2019,babarit_theoretical_2015}, before presenting the results from the simulations. We finish with a discussion on future work and conclusions.

\section{Lattices}

We differentiate between arrays and lattices as follows: an array is a realization of a geometric lattice consisting of an arrangement of cylindrical free-surface-piercing scatterers.

\begin{figure}
    
    \includegraphics[width=1\textwidth]{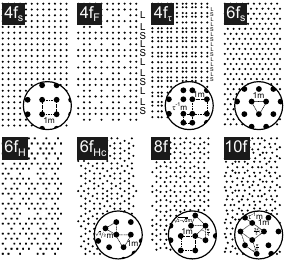}
    \caption{Plan views of the waveguide designs and notation used in this work.  Descriptions of the lattice geometries can be found in the text. The waveguides are plotted to scale: the lattice parameter of \Ffs{} is 1 m and the diameter of each element is 0.4 m. The arrays vary slightly in size, with the largest, \Ffs, measuring 11.4 m $\times$ 21.4 m. Each waveguide is placed at the centre of a 40 m $\times$ 40 m computational domain with waves impinging from the left. The insets show geometric details. For lattices \FfF{} and \Fft{} the Fibonacci sequence is indicated via the use of L and S segments.}
    \label{lattices}
\end{figure}

Our interest lies in exploring the wave propagation properties of arrays with varying rotational symmetries and periodic/quasiperiodic structures.  The designs investigated are shown in Figure \ref{lattices}, encompassing  archetypes and simple modifications thereof of 4--, 6--, 8--, and 10--fold rotational symmetries, where they are generally referred to by their $n$-fold symmetry as $n$f. The finer details of their structure and nomenclature are discussed below. In an effort to study the effects of geometry independently of other characteristics of the array, where possible, the most common nearest neighbour distance is fixed at $NN = 1$ m.  A cylinder is placed at each lattice point (section \ref{sec:methods}). The filling fraction (FF) is defined as the area occupied by cylinders divided by the total area of the array (see Figure \ref{fig1}(b)). Each infinite lattice has a calculable FF. However, in keeping with the desire to deal only with properties of finite arrays, here FF is calculated directly by summing the cylinder cross-sectional areas and dividing by the area of the bounding box of the array. According to the work of Hu \emph{et al.} \cite{hu2005refraction}, the refractive index is then $n=\sqrt{1+FF}$. Number of elements, FF and $n$ are given for each array in Table \ref{latticesT}.  For the purposes of comparison both periodic and aperiodic arrays are investigated. We note that the derivation of Hu \emph{et al.} contains no reference to the array geometry at any stage, implying its validity for both periodic and aperiodic arrays.  We also note that it is derived in the long wavelength limit; however, the correspondence principle implies that long-wavelength-regime phenomena will also apply in the Bragg regime we are operating in, although they will be dominated by other effects.

\subsection{Periodic arrays}

The two simplest classes of periodic lattice are 4--fold (square) and 6--fold (hexagonal), arrays based on which are shown in Figure \ref{lattices} and labelled as \Ffs{} and \Sfs{} respectively, where the subscript s denotes `simple'. All continuous lattices in two-dimensional space must have even-number $n$, due to their indistinguishability under rotational inversion \cite{lifshitz1996symmetry}.

\subsection{Quasiperiodic arrays}

The remaining arrays are quasiperiodic. Here, we discuss the structure and characteristics of their underlying lattices.

\begin{itemize}[leftmargin=*]
  \item Missing-element square Fibonacci lattice
  \begin{itemize}[leftmargin=*, label={}]
  \item Array \FfF{} is based on a square lattice with elements missing in an quasiperiodic fashion.  The missing elements are chosen based on the binary Fibonacci sequence (\textit{LSLLSLS..}), where the individual elements of the sequence are either \textit{a} (\textit{S}, short) or 2\textit{a} (\textit{L}, long). The sequence is indicated in the figure. 
  \end{itemize}

 \item $\tau$-scaled square Fibonacci lattice
   \begin{itemize}[leftmargin=*, label={}]
   \item Array \Fft~is a square lattice of elements placed according to 2 perpendicular Fibonacci sequences, one of which is indicated in the figure, where $L/S$ is equal to the golden ratio $\tau$. As there are $\tau$ times as many $L$ segments as $S$ segments in each (infinite) sequence, the $NN$ in this case is $1/\tau=0.618..$, and the FF is accordingly high \cite{lifshitz2002square}.
   \end{itemize}

  \item Hexagonal lattices based on the $H_{00}$ lattice
  \begin{itemize}[leftmargin=*, label={}]
  \item Array \SfH{} is based on a periodic hexagonal lattice with elements removed to produce a quasiperiodic arrangement, similar to \FfF{}.  The lattice points decorate the vertices of the $H_{00}$ lattice identified by Coates \emph{et al.} \cite{Coates2023hexagonal}, where the short and long edges of the rhombic and hexagonal tiles are both set to 1 m \cite{Coates2024hexagonal}. Array \SfHc{} is also produced using the $H_{00}$ lattice with tile edges set to 1 m, but instead uses the centre points of the tiles. In these cases, the relationship to the Fibonacci sequence is not trivial, so it is not indicated. Further details can be found in refs. \cite{Coates2023hexagonal,Coates2024hexagonal}.
  \end{itemize}

  \item Ammann-Beenker lattice
  \begin{itemize}[leftmargin=*, label={}]
  \item Array 8f is constructed from the vertices of the octagonal Ammann-Beenker lattice \cite{Beenker1982algebraic,Socolar1989simple,arnouxtilings}, which consists of squares and rhombi with an internal angle of $\frac{\pi}{2}$~rad. The edge length of squares and rhombi in the lattice is 1 m, but elements are separated by $\sqrt{2-\sqrt{2}}$~m along the short diagonal of the rhombi.
\end{itemize}

 \item Penrose lattice
  \begin{itemize}[leftmargin=*, label={}]
  \item Array 10f is formed from the vertices of a Penrose P3 lattice \cite{penrose1979pentaplexity}, which is perhaps the most well-known quasiperiodic geometry. The lattice is composed of two types of rhombi, with internal angles $\frac{\pi}{5}$~rad (thin) and $\frac{2\pi}{5}$~rad (fat), arranged according to matching rules. Though it is often described as \textit{pentagonal}, its indistinguishability under rotational inversion gives this lattice overall 10--fold symmetry \cite{lifshitz1996symmetry}. The rhombus edge length is chosen to be 1 m. This again leads to elements being separated by less than 1 m, here, along the short diagonal of the thin rhombi, with the separation being equal to $\tau^{-1}$~m.
  \end{itemize}
    \end{itemize}

\begin{table}
    \centering
    \begin{tabular}{cccc}
         Array&  $N$ elements &  Filling fraction (FF) & Refractive index $n$\\
         \hline
         \hline
         
         \Ffs{} &  264 & 0.136 & 1.066\\
         \FfF{} &  199 & 0.118 & 1.057\\
         \Fft{} &  288 & 0.184 & 1.088\\
         \Sfs{} & 250 & 0.150 & 1.072\\
         \SfH{} & 202 & 0.110 & 1.054\\
         \SfHc{} & 232 & 0.138 & 1.067\\
         8f & 240 & 0.153 & 1.074\\
         10f & 246 & 0.154 & 1.074\\

    \end{tabular}
    \caption{Number of elements, filling fraction and refractive indices of the arrays.}
    \label{latticesT}
\end{table}

\begin{figure}
\includegraphics[width=1\linewidth]{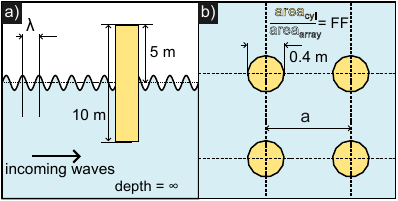}
    \caption{\textbf{(a)} Schematic of the simulation setup. Cylinders with a length of 10 m are submerged 5 m into water, which is set to infinite depth. \textbf{(b)} Top-down view of the cylinders in the square array, which demonstrates how the filling factor (FF) is calculated.}
    \label{fig1}
\end{figure}

\section{Methods}

\subsection{Numerical domain}\label{sec:domain}
The array is placed at the centre of a 40 m $\times$ 40 m square grid, with a grid resolution of 512 $\times$ 512 and a resultant cell size of 7.8 $\times$ 7.8 cm.  The calculations result in linear solutions to a linear problem.  This means that the grid resolution of the water surface does not affect the calculation or the results, in contrast with the effects of the cell size in FNLPF simulations.  The linear solution is sampled at each point in the grid, and the grid size set as the lowest power of 2 that preserves the smallest details observed in the data.  Convergence testing of the grid resolution for 8 values in the range 64 to 1024 yields a constant (L2 error norm vs $\Delta x$) gradient of 0.175, where $\Delta x$ is equal to the computational domain edge length (40 m) divided by the grid resolution. For well-converged results, the gradient is not greater than 1.  The gradient in our test is much smaller than 1, so the results are converged.  The gradient in this case is entirely due to the improvement expected from an increase in sampling resolution, and cannot therefore decrease further.

The waves are incident from the left hand side. They impinge on the array, are scattered, and we observe the steady state response of the water waves in the domain.  The water is set to infinite depth so that the waves can be considered deep water waves.  The arrays themselves are composed of fixed 0.2 m radius, 10 m long, 13-gon prism approximations to cylinders, with 5 m submerged. The number of sides is chosen to maximise resolution whilst keeping within memory constraints, and also to avoid any rotational symmetries in common with the arrays. As noted, the most common separation present in the array is set to 1 m. Wave heights are scaled from unity in Capytaine: any wave height or motion amplitude can be retrieved by multiplying the result by the desired value \cite{ancellin_capytaine_2019}, and the results are valid in the linear regime.

\subsection{Water wave modelling with Capytaine}\label{sec:methods}
The interaction between the  waves and the arrays is simulated herein using the Capytaine open-source software (Ancellin and Dias \cite{ancellin_capytaine_2019} and Babarit and Delhommeau \cite{babarit_theoretical_2015}).  Capytaine is a Python-based boundary element method (BEM) solver for linear potential flow in water waves \cite{capysc}. Based on previous code known as NEMOH \cite{babarit_theoretical_2015}, the linear potential flow approach has been widely used including multi-frequency and multi-direction wave loads on wind turbine platforms \cite{kurnia2022computation}.  Starting from the assumptions of inviscid, irrotational and incompressible flow, the linear potential flow theory solves the problem in the frequency domain and is able to predict the radiation and diffraction processes.  As a potential flow solver, the approach can be used to predict waves until they become near nonlinear and eventually break, at which point alternative (and computationally more expensive) approaches are required, such as full Navier-Stokes solvers (\cite{OpenFOAM} and \cite{Dominguez_etal_2021}). The predictions herein are \textit{only} linear, so as the frequency increases, the maximum wave height for which they are applicable decreases. According to the limit for strict linearity $\nicefrac{H\omega^2}{4g\pi^2}<0.001$, our waves are perfectly modelled only for a height $H$ between 4.3 cm at 3.0 rad s $^{-1}$ and 6 mm at 8.0 rad s $^{-1}$.

The calculations were performed on the Jeremiah Horrocks Institute High Throughput Cluster, which comprises 13 Dell R650XS nodes each with 48 physical cores and 256 GB RAM running Oracle Linux 8.9 and Slurm 20.11.9.

The arrays are constructed using Capytaine's internal routines.  One cylinder is placed, then duplicated to create the array.  Tridecagonal prisms are used as they maximise usage of available RAM.

\begin{figure}
    \centering
    \includegraphics[width=1\linewidth]{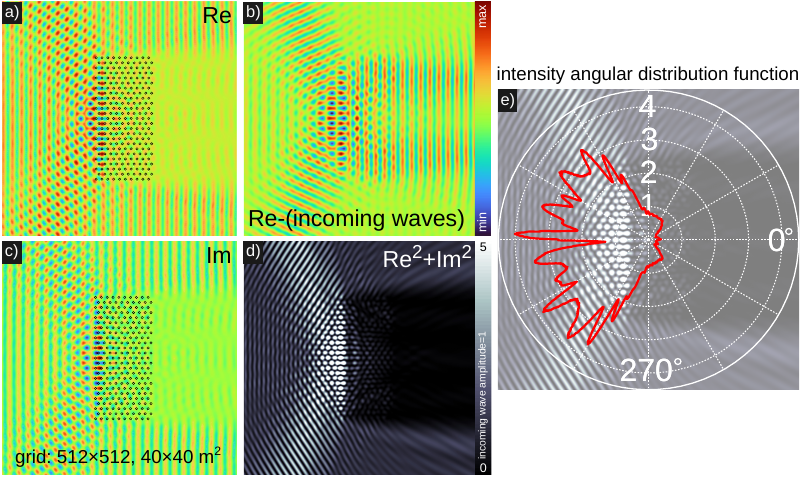}
    \caption{Example output from Capytaine, processed with MATLAB. Hexagonal array, $\nicefrac{ka}{\pi}$ = 1.288. \textit{a,c):} The real and imaginary components of the solution to the wave equation, with the array superimposed. \textit{b):} The real component with the incoming waves subtracted to aid identification of diffracted waves. The colourbar indicates that each plot is normalized to its own maximum and minimum values. \textit{d):} intensity, defined as the sum of the squared components. For equal intensity to the incoming waves, this has unit value $(1^2+0^2=1)$. \textit{e):} the angular distribution function of the intensity, superimposed on the intensity map. Each point on the polar curve is the mean value of a profile plotted from the centre to the perimeter of the intensity map.}
    \label{sample}
\end{figure}

Capytaine can provide many kinds of output. Here we use the water free surface, provided as separate grids of the real and imaginary components of the solution to the wave equation.  We use MATLAB to present and perform calculations on these datasets.  An example dataset is shown in Figure \ref{sample}, showing the interaction between waves and the \Sfs~array. The two components of the free surface are shown in panels (a,c). All data in this manuscript presented with this colourmap are normalized to the range of the individual dataset.

We generate a reference wavefield of appropriate frequency and phase to represent the incoming waves and subtract it from the real component to give panel (b).  This treatment aids in identification of scattered waves, in particular their direction. However, if there is any significant interaction between the waves and the array (i.e., if a phase difference is introduced, and/or the amplitude is altered), the area to the right of the array is dominated by the subtracted wavefield. 

In panel (d) we add the real and imaginary components squared, which gives the intensity of the wave energy. 
Data presented with this colourmap always represent intensity, and are normalized to the range 0-5, with 1 being the same intensity as the incoming waves.

In panel (e), a polar plot shows the angular distribution function (ADF). This function is generated by taking a profile from the centre to the periphery of the intensity map, at the denoted angle, and plotting the mean value of this profile as a point on the red curve. For profiles of either the real or imaginary components, it would be appropriate to use the root mean square; as we use profiles of the intensity, there is no inherent periodicity and so the mean is appropriate. The influence of the varying length of the profiles in the non-circular data is minimized by use of the mean rather than the integral.  In the example given, this shows strong intensity in the directions of scattered waves, and zero intensity in the direction of propagation after interaction with the array. The ADF calculation includes the area inside the array.  

This system has mirror symmetry about the horizontal axis. The apparent asymmetry in the results, particularly noticeable in the ADF, is due to aliasing between the array and the simulation cell grid, and would be reduced for a higher-density grid. Another non-mirror-symmetric element, shown in Figure \ref{fig1}, is the 13-gon prism used to represent a cylinder, though the effect of this is negligible.

\section{Results and discussion}

There are several facets to the interaction between the array waveguides and the impinging water surface waves.  We begin with a description of the `blocking' behaviour of the waveguides; that is, the capacity of a waveguide to prevent wave energy from reaching the leeward side of the array.  This behaviour depends, to varying degrees according to the various waveguides, on the primary Bragg resonance.

We then describe other Bragg diffraction behaviour. As we are dealing with three-dimensional simulations of 2-dimensional arrays, we have the opportunity to observe non-primary Bragg resonances characteristic of the rotational symmetry of the waveguides. These cause effective redirection of the wave energy.

We finish with a description of the transmission of wave energy through the waveguides, which is accompanied and enabled by the establishment of a periodic relationship between wave and waveguide.

\subsection{Blocking}

\begin{figure}
    \centering
    \includegraphics[width=0.8\textwidth]{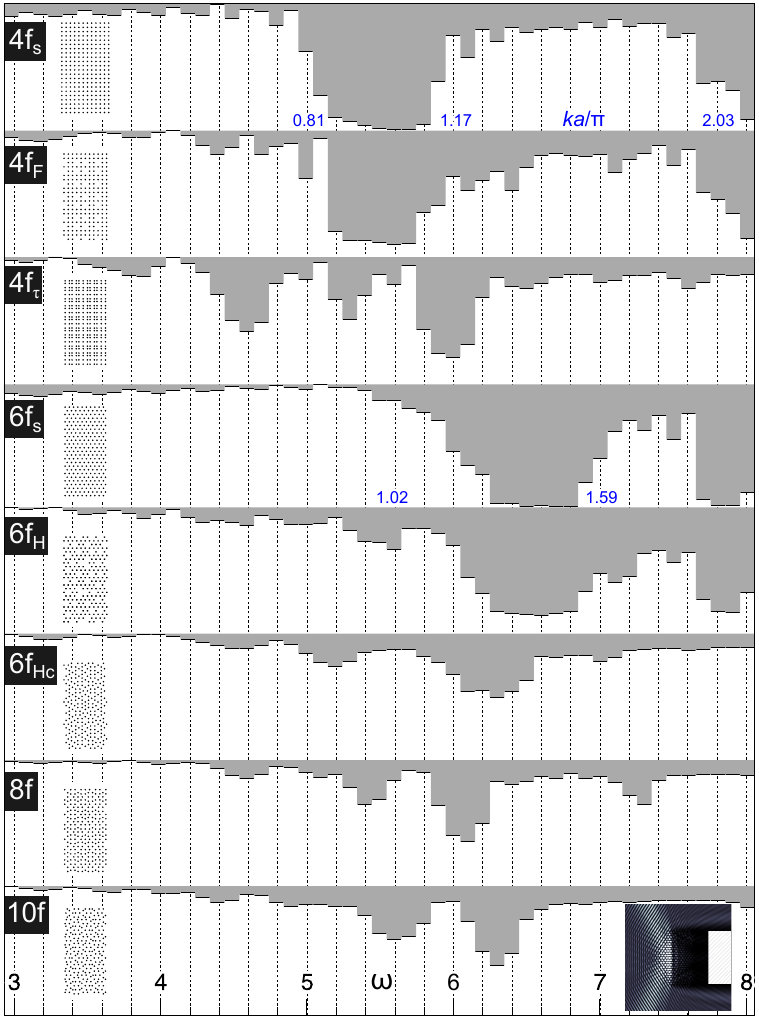}
    \caption{Blocking curves for the arrays with monochromatic waves incident from the left, against $\omega$. The curves represent the mean value in the crosshatched area of the intensity graph shown in the lower-right corner. The shaded region represents the blocked waves. Each curve is plotted with a unitary $y$-scale. The $\omega$ $x$-scale is for all curves.  For curves obtained from an essentially periodic array, relevant values of $\nicefrac{ka}{\pi}(=\nicefrac{\omega^2}{\pi g})$ are provided (blue).}
    \label{blocking}
\end{figure}

The capacity of an array to remove intensity from an area behind the array versus the angular frequency of the incoming waves $\omega$, was measured to give `blocking' curves, which are presented in Figure \ref{blocking} for each array. \textit{Blocking dips} correlate with complete or incomplete band gaps. Using the ADF described above to compile this data presents two problems: firstly, it is strongly directionally selective; secondly, its calculation includes data inside the array. These problems limit its applicability to the question of how effective a given array is at blocking waves, so we take another approach.  In the inset in Figure \ref{blocking}, we indicate a region of interest (ROI) to the right of the array. All datasets in the figure are generated using this ROI, with the mean value of intensity (as defined above) within the ROI used to represent the capacity of an array to remove intensity from the ROI.

A numerical summary of these results is compiled in Table \ref{btable}, where we show the maximum blocking \% for any frequency, and the total blocking observed over the full frequency range.  Earlier work by Ha \emph{et al.} using the multiple-scattering method -- in which diffracted waves are linked to the incoming wave and represented by Fourier–Bessel expansions to derive the band structure of similar arrays -- shows the existence of complete Bloch band gaps at $\nicefrac{ka}{\pi}=1.2$ for a square array and $\nicefrac{ka}{\pi}=1.6$ for a hexagonal array \cite{ha2002propagation}. Our results show almost 100\% blocking dips starting at lower frequencies and bounded by these values at the upper end.

These results indicate that a specified periodic array can constitute an effective strategy for blocking water wave propagation for a given range of angular frequencies.

In contrast, arrays \FfF~and \SfH, which are identical to \Ffs{} and \Sfs{}, but with elements removed in a quasiperiodic fashion, have slightly lower blocking in their respective Bloch band gaps. However, they do show improved blocking in regions away from the major blocking dips. This leads to overall better blocking performance over the entire frequency range, as tabulated in Table \ref{btable}.

Previous analytical approaches to two-dimensional problems have shown that the existence of zero-reflection wave modes is expected if and only if the scattering potential is symmetrical with respect to the wave propagation direction \cite{xie2011analytical,goyal2025optimal}. The behaviour of two-dimensional wave-obstacle systems cannot be meaningfully extrapolated to fully describe three-dimensional interactions between water waves and an array waveguide.  However, the lifting of symmetry via the addition of vacancy defects will contribute to the removal of zero-reflection modes, and therefore the increased blocking outside of the band gap.  This loss of symmetry also disrupts the primary Bragg resonance, resulting in some transmission of frequencies inside the band gap.

A hexagonal lattice is denser than a square lattice of the same lattice parameter.  Therefore, of the periodic arrays, \Sfs{} has the highest filling fraction, which seems a reasonable explanation of the slightly better blocking performance of the hexagonal arrays.  When this is combined with symmetry removal via the addition of vacancy defects, the combination of low filling fraction and optimal blocking performance is achieved.

For the quasiperiodic arrays based on $\tau$, or on higher rotational symmetries,  the blocking curves are characterized by more dips, that individually are narrower than those for the periodic arrays.  The behaviour of \Fft~is strikingly similar to 8f, in terms of the locations of the blocking dips. The action of the arrays over a wide frequency range indicates that quasiperiodic lattices could contribute to broadband blocking strategies in addition to their rotational isotropy (which we do not investigate here).  The quasiperiodic arrays cannot by definition be perfectly periodic, or symmetrical, in the propagation direction, which limits the possibilities for both zero-reflection modes and a Bloch band gap.

The overall relative behaviour of the arrays according to their geometry is analogous to the situation of diffraction from crystals.  Quasicrystals, due to their quasiperiodicity, have a diffraction pattern of infinite density. In practice, most of the diffraction peaks are too dim to observe, so a discrete diffraction pattern is observed. This pattern is of lower intensity than that from a periodic crystal, in which the scattering planes contributing to a particular low-index peak are far more numerous than those for a quasicrystal. Our quasiperiodic arrays \Fft, \SfHc, 8f and 10f show analogous behaviour in their multiple blocking dips which each are smaller than those for the periodic arrays.  This behaviour is reminiscent of the fractal transmission structure exhibited by quasicrystalline metamaterials \cite{zolla1998remarkable,davies2023graded}.

The difference between the \SfH~and \SfHc~blocking curves is notable, as they are based on the same underlying lattice but with a different basis location.  There therefore seem to be at least two independent components to the blocking behaviour: one dependent on standing waves in a periodic lattice, after the Bloch theorem, and one dependent on the underlying quasiperiodic ordering. In the blocking from \SfH{} the curve is dominated by the component from the periodic array, whereas in the blocking from \SfHc{}, this component is entirely absent, revealing a blocking curve with the same characteristics as the other quasiperiodic arrays.

\begin{table}
    \centering
    \begin{tabular}{cccc}
         &  Maximum blocking & Maximum &  Total blocking over\\
        Array &per frequency (\%) & $\omega$ (rad s$^{-1}) [\nicefrac{ka}{\pi}]$  &  frequency range (\%)\\
         \hline
         \hline
         
         \Ffs &  99.2& 5.6 [1.02] & 31.6\\
         \FfF &  97.6& 5.6 [1.02] & 32.9\\
         \Fft &  84.2& 6.0 & 23.4\\
         \Sfs &  98.8& 6.8 [1.50] & 33.4\\
         \SfH &  96.4& 6.6 [1.41] & 36.5\\
         \SfHc &  54.8& 6.3 & 15.6\\
         8f &  70.1& 6.1 & 15.4\\
         10f &  68.0& 6.3 & 16.7\\
    \end{tabular}
    \caption{Blocking performance of arrays.}
    \label{btable}
\end{table}

\subsection{Bragg diffraction}

In x-ray and electron crystallography, a beam of x-rays or electrons is made incident on a crystal. The scatterers (atoms) in a crystal can be grouped into families of parallel planes of atoms, where one is differentiated from another via its Miller indices.  The Miller indices are the numbers of unit cells in each direction to define a family of planes via its normal. For example, (001) in a cubic crystal involves moving 0 unit cells in $x$, 0 unit cells in $y$ and 1 unit cell in $z$; this defines the normal to, and thus refers to, the $xy$-planes [(010) refers to the $xz$-planes, (100) refers to the $yz$-planes].  Each unique family of planes has a unique set of Miller indices, and produces a family of harmonic peaks in the diffraction pattern. The angle at which a peak is located is given by the Bragg law $n\lambda = 2d\sin{\theta}$, where $d$ is the separation between planes.

In a two-dimensional array of scatterers, the planes are now lines of scatterers, with particular separations. In general, all of the arrays produce strong diffraction when the Bragg condition is satisfied by one or more separations in the array.  The diffracted beams are directional and diffuse, which is consistent with the small number of scatterers in the arrays\cite{liu2019bloch}.

\begin{figure}
    \centering
    \vspace{-0cm}
    \includegraphics[width=1\linewidth]{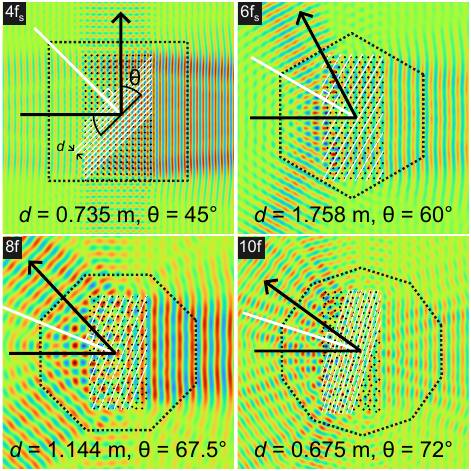}
    \caption{2-dimensional Bragg diffraction from a subset of the arrays. Overlaid in dotted lines are the high-symmetry $n$-gons associated with the rotational symmetries of the arrays. Black arrows indicate the incoming wave direction and angle associated with the rotational symmetry $n$ of the array ($\pi/2 - \pi/n$). Variables $d$ and $\theta$ correspond to those in the Bragg law. The white lines are a visual representation of the line family reconstructed from the Bragg law without knowledge of the scattering array. The data presented is the real component (see Section ) with incoming waves subtracted.}
    \label{bragg}
\end{figure}
    
Here, we use the Bragg law 
to extract the apparent interlinear distance $d$ of the arrays when the wavefront is normal to the array. For infinite periodic arrays $d$ is a simple function of $a$, the lattice parameter. For finite periodic arrays, it is a function of $a$ with an error due to the finite radius of the scatterers.

For quasiperiodic arrays, there is no single lattice parameter, though examination of the arrays reveals that their geometries depend on arrangements of two characteristic separations.  For example, in the Penrose lattice 10f, these separations are 1 m, the rhombus edge length, and $\tau^{-1}$~m, the width of narrow rhombi. Diffraction peaks then arise from every possible combination of these distances, with relative intensities given by the relative density of occurrences of each combination. For more discussion on this topic, the reader is referred to e.g. \cite{diehl2003low,lifshitz2002square}.

All the arrays have been simulated over the range $3.0<\omega<8.0$.  The procedure we have used in Figure \ref{bragg} is:

\begin{enumerate}
    \item Identify the $n$-fold symmetry of the array
    \item Select the value of $\omega$ that produces diffraction at the expected angle (i.e., through the side of the $n$-gon adjacent to the side impinged upon by incoming waves)
    \item Calculate $d$ from known $\theta$ and $\omega$ using the Bragg law
\end{enumerate}

Close to the Bragg angles for a given lattice, constructive interference may occur for a range of frequencies, with the diffracted beam sweeping a concomitant range of angles.  Therefore, it is appropriate to use the Bragg law to extract $d$ only when a lattice has one or more easily identified directions of symmetry. Figure \ref{bragg} shows this analysis for the \Ffs{}, \Sfs{}, 8f, and 10f arrays, where a black arrow indicates the incoming and diffracted wave direction and angle associated with the rotational symmetry $n_R$ of the array ($\nicefrac{\pi}{2} - \nicefrac{\pi}{n_R}$). For the simple square array, the 2D Miller indices of the line family are (11), and $d$ is found to be close to the expected value of $\nicefrac{\sqrt{2}}{2}$. For the simple hexagonal array, the lines are in the 2d hcp (100) family. Adjacent lines in this family have $d=\nicefrac{\sqrt{3}}{2}$; we observe close to twice this value, as this is the Bragg reflection available in our frequency range. The deviation from the expected values is consistent with the finite size of the array and the finite radius of the scatterers.  For example, for array \Ffs, scattering is evident from a continuous range of lattices between those defined by the innermost points and outermost points of the cylinders in the array. In any finite periodic array, the number of elements and the dimensions of the array lead to a minimum total number of lines ($N_\text{lines}$) belonging to any set of Miller indices; in array \Ffs, at an angle of $\nicefrac{\pi}{4}$~rad, (Miller indices (11)), $N_\text{lines}$ is 11. This gives an approximate $\pm5$\% error from the limits of $d$, given by ($N_\text{lines}\pm\nicefrac{4R_\text{cylinder}}{\sqrt{2}})\div N_\text{lines}$, where $R_\text{cylinder}$ is the cylinder radius.

The \FfF{}, \SfH{}, and \SfHc{} arrays show the same broad behaviour as the simple structures with this analysis. The 8f and 10f arrays have well-defined scattering angles, which permits the extraction of an apparent `interlinear' distance, shown in the figure as $d$.

The $\tau$-scaled 4--fold array \Fft, not shown here, has ambiguous behaviour: although it has 4--fold symmetry, it never strongly diffracts from a set of lines at an angle of $\nicefrac{\pi}{4}$~rad as expected. However, weak diffraction at multiple wavelengths yields values for $d$ of 1.00 m and 1.39 m.

\subsection{Lattice coherences}

In our data, transmission of surface wave energy through an array (visualised as intensity downstream of the array) is usually observed simultaneously with a periodic variation in intensity inside the array. The period is expressed via a simple ratio between the array periodicity and the incident surface wavelength. We call this periodic variation in intensity a \textit{lattice coherence} (LC).

Various LCs were observed across the arrays, corresponding with effective transmission. In each case, the wavelength is distorted from the incident wavelength to fit the LC.  The pattern of intensity within the array is examined and the period and hence unit cell determined via feature correlation. Following this, the number of wave crests in the unit cell of the LC is counted. The number of lattice periods is readily countable by superimposing the array on the data.

\begin{figure*}
    \centering
    \includegraphics[width=0.75\linewidth]{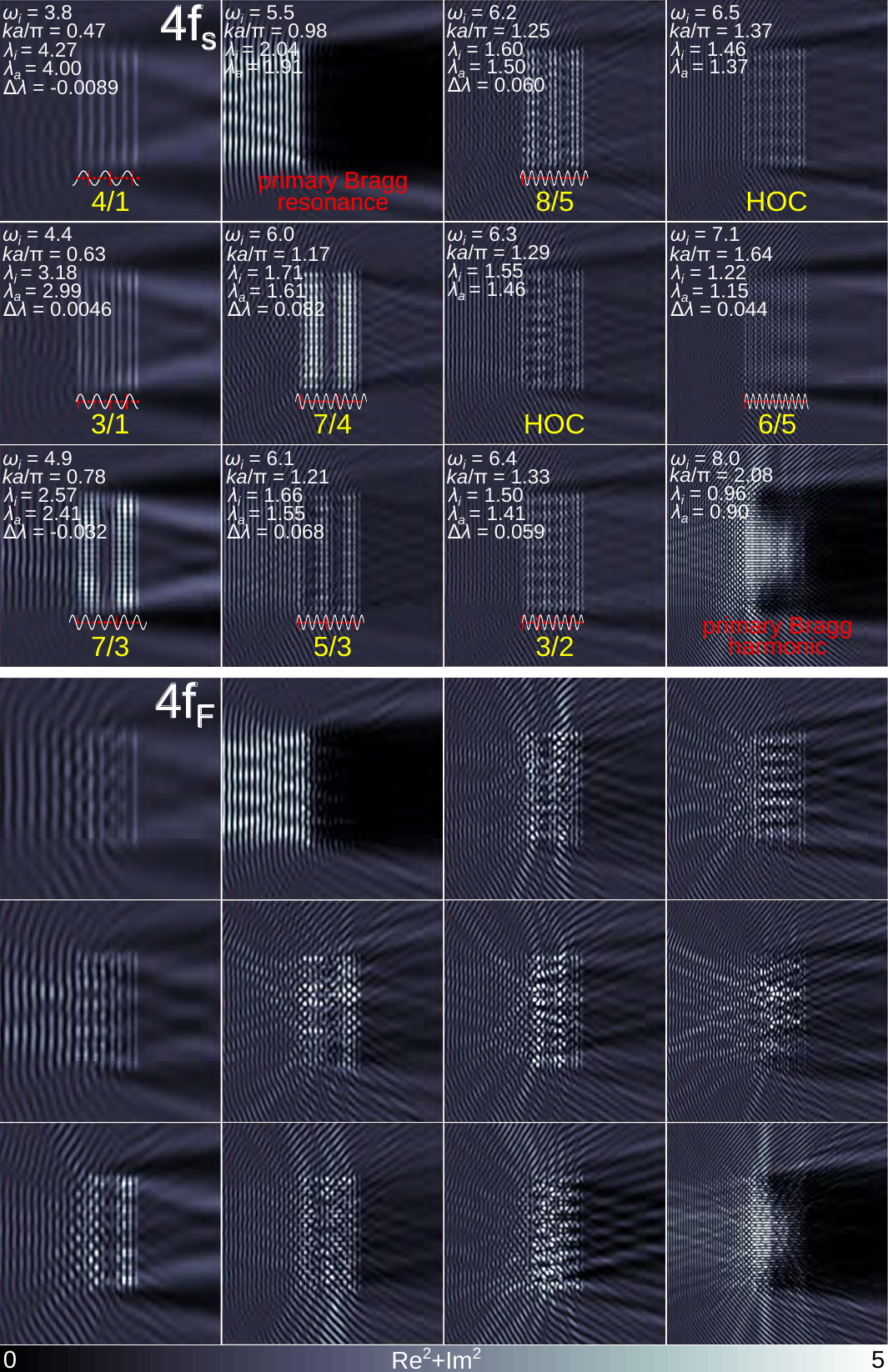}
    \caption{\textit{Upper half}: LCs and reflection in array \Ffs. Plots are of intensity, normalized to the range indicated. Further discussion is in text. \textit{Lower half}: the corresponding intensity plots for array \FfF, showing disruption of LCs.}
    \label{sqres}
\end{figure*}

In Figure \ref{sqres}, we explore the intensity maps for the \Ffs{} array, the simplest case, and the structurally similar \FfF{} array, for certain values of $\omega$. The angular frequency $\omega$ is an input variable, and runs from 3.0 -- 8.0 rad s$^{-1}$ in increments of 0.1. In each panel for the \Ffs{} array, we list certain values:

\begin{itemize}
    \item
    $\omega_i$: the angular frequency of the incident surface wave (rad s$^{-1}$)
    \item
    $\nicefrac{ka}{\pi}$: a wave-lattice structure interaction parameter, provided for comparison to the work of Ha \emph{et al.}\cite{ha2002propagation}
    \item
    $\lambda_i$: the incident surface wavelength, calculated from $\lambda_i=\frac{2\pi g}{\omega^2}$
    \item
    $\lambda_a$: the wavelength of the wave inside the array adjusted for the lattice index of refraction, given by $\lambda_a=\nicefrac{\lambda_i}{n}$, where $n$ is the refractive index calculated from the filling fraction of the array $n=\sqrt{1+FF}$~\cite{ha2002propagation}
    \item
    $\Delta \lambda$: the fractional difference between $\lambda_a$ and the LC ratio, or wavelength inside the array, $\Delta\lambda=1-\lambda_a\div\nicefrac{N_p}{N_\lambda}$; this is a measure of the distortion the waves undergo to reach LC
    \item
    LC ratio (at the bottom of each relevant panel): the ratio of number of lattice periods ($N_p$) to number of wavelengths ($N_\lambda$); this is numerically equal to the wavelength of the waves inside the array, given that the lattice period is 1 m.
    
\end{itemize}

Identifiable LCs are indicated; these follow a notation in which the numerator is the number of lattice parameters and the denominator is the number of wavelengths.  A graphical representation of the LC is also shown, showing the relationship between the surface waves (white) and the array periodicity (red).


We borrow a phrase from condensed matter physics: \textit{higher order commensurate} (HOC), to describe the situation where a LC is visible via a periodic pattern of intensity, but where a complete cycle of the LC does not fit inside the array.

For array \Ffs, LCs are strongly correlated with transmission, and, to achieve LC, the wavelength inside the array may be distorted by up to $\Delta\lambda=8.2\%$ in our data. A LC results in a rational relationship between wave and array: waves at the same position in each unit cell of a LC encounter a scatterer at the same angle of phase. Each unit cell is therefore symmetrical; the transmission function for each unit cell therefore has zero-reflection modes, and the stacking of unit cells across the array extends the zero-reflection modes across the array.

The large degree of distortion achieved to gain LC is strongly suggestive that the waves will follow such a LC even if the array undergoes a moderate transformation. We leave this possibility for future work.

The corresponding intensity maps for array \FfF, based on a square lattice with elements removed in a Fibonacci sequence, are shown in the lower half of Figure \ref{sqres} for comparison.  In analogy to the scattering of electron waves by defects in periodic crystals \cite{bloch1929quantenmechanik}, the LCs observed in \Ffs~are strongly disrupted by the defects in \FfF, resulting in reduced transmission.  The mechanism for this disruption is the removal of symmetry and periodicity, and therefore zero-reflection modes.

\section{Conclusions}

We have used the open-source Capytaine software to investigate the behaviour of water surface wave array waveguides for a range of array geometries, with the number of array elements ranging from 199-288. Using this approach, we have successfully replicated an earlier result that finds band gaps for some of these array types \cite{ha2002propagation}, and indeed have shown the existence of band gaps for all the arrays, strongly indicating that this is a viable strategy for blocking and/or reflecting wave energy. Our methodology allows us to directly observe many wave phenomena in real space, for example Bragg diffraction, refraction and resonance.  Significantly, our use of simple archetypes of periodic and quasiperiodic lattices demonstrates the potential of this approach to investigate any kind of array waveguide within the linear water wave regime.

A particularly striking result is the degree of reflection that is achieved by many of these geometries. Nearly 100\% reflection is observed in several cases, strongly supporting the adoption of such arrangements in coastal defence strategies.

Of the arrays tested, the periodic arrays and two quasiperiodic arrays are characterised by a single lattice parameter.  The blocking curves of these arrays are characterised by two large blocking dips caused by the primary Bragg resonance and formation of a Bloch band gap.

The quasiperiodic arrays based on the golden ratio $\tau$ or on higher-order rotational symmetries generated blocking curves characterised by several blocking dips that are smaller than those for periodic arrays.

The array with the most effective blocking over the frequency range is hexagonal with quasiperiodically located vacancy defects. This array also has the lowest filling fraction of those studied.

Transmission through the waveguides is heavily influenced by relationships between the wavelength and the array geometry, here called lattice coherences (LC). These LCs can be easily disrupted by, e.g., removing array elements, further enhancing the ability of array waveguides to block wave propagation. Here, we have placed vacancy defects in a systematic quasiperiodic fashion. Future work could explore different geometries of vacancies, for example, periodic or random.

Beyond this, the observation, from simple inspection of Figure \ref{sqres}, that the waves are in some kind of LC for a greater range of frequencies than is blocked by band gaps, strongly suggests that it may be more effective to control waves by using the transformation-optics-inspired approach of modifying the arrays to aid propagation in preferred directions.

In this work, we have normalised the arrays by using the same array element dimensions.  To extract the detailed behaviour of geometry in isolation from such confounding factors as array density, it may be instructive to perform the study normalised to other factors, such as the array refractive index, which can be modified by changing element radius and thus filling fraction.

The use of full non-linear potential flow calculations in future applications of this approach would allow investigations of energy transfer between different modes.

Finally, the results highlight the need for experimental investigations of array waveguides.  Here, we probe only the linear regime, and find significant opportunities. When the full range of water surface wave phenomena are allowed to interact with array waveguides, additional avenues for research may become apparent.

\section{Acknowledgments}

JAS would like to acknowledge useful discussions with P. Moriarty, R. Lifshitz, M. Ancellin and B. A. Patterson.  The figures in this manuscript were produced using Inkscape, Plot2 and Veusz. The data were analysed using MATLAB and Fiji (ImageJ). 

Funding: SC was supported by EPSRC grant EP/X011984/1. The computational work was supported by the Jeremiah Horrocks Institute.

 \bibliographystyle{elsarticle-num} 
 \bibliography{bib}

\end{document}